\documentclass[conference]{IEEEtran}

\IEEEoverridecommandlockouts

\usepackage{cite}
\usepackage{amsmath,amssymb,amsfonts}
\usepackage{algorithmic}
\usepackage{graphicx}
\usepackage{subfigure}
\usepackage{textcomp}
\usepackage{xcolor}
    
\usepackage{amsthm}
\usepackage[ruled]{algorithm2e}
\usepackage{mathrsfs}
\usepackage{bm,epsfig,amsthm,url}
\usepackage{indentfirst}
\usepackage{epstopdf}
\usepackage[top=0.75in, left=0.7in, bottom=1.1in, right=0.7in]{geometry}

\theoremstyle{definition}
\newtheorem{lemma}{Lemma}

\def\BibTeX{{\rm B\kern-.05em{\sc i\kern-.025em b}\kern-.08em
    T\kern-.1667em\lower.7ex\hbox{E}\kern-.125emX}}
\begin{document}

\title{Stochastic Beamforming for Reconfigurable Intelligent Surface Aided Over-the-Air Computation
\thanks{This work is supported by the National Natural Science Foundation of China (NSFC) under grant 61932014.}}

\author{
  \IEEEauthorblockN{$\text{Wenzhi Fang}^{\ast \dagger \S}$, $\text{Min Fu}^{\ast \dagger \S}$, $\text{Kunlun Wang}^{\ast}$, $\text{Yuanming Shi}^{\ast}$, and $\text{Yong Zhou}^{\ast}$
  }\\
 \IEEEauthorblockA{
  $^{\ast}$School of Information Science and Technology, ShanghaiTech University, Shanghai 201210, China \\ 
  $^{\dagger}$Shanghai Institute of Microsystem and Information Technology, Chinese Academy of Sciences, China\\
  $^{\S}$University of Chinese Academy of Sciences, Beijing 100049, China\\
  Email: \{fangwzh1,fumin, wangkl2, shiym, zhouyong\}@shanghaitech.edu.cn}}

\maketitle

\begin{abstract}
Over-the-air computation (AirComp) is a promising technology that is capable of achieving fast data aggregation in Internet of Things (IoT) networks. 
The mean-squared error (MSE) performance of AirComp is bottlenecked by the unfavorable channel conditions. This limitation can be mitigated by deploying a reconfigurable intelligent surface (RIS),  which reconfigures the propagation environment to facilitate the receiving power equalization. 
The achievable performance of RIS relies on the availability of accurate channel state information (CSI), which however is generally difficult to be obtained. 
In this paper, we consider an RIS-aided AirComp IoT network, where an access point (AP) aggregates sensing data from distributed devices. 
Without assuming any prior knowledge on the underlying channel distribution, we formulate a stochastic optimization problem to maximize the probability that the MSE is below a certain threshold. 
The formulated problem turns out to be non-convex and highly intractable. 
To this end, we propose a data-driven approach to jointly optimize the receive beamforming vector at the AP and the phase-shift vector at the RIS based on historical channel realizations. 
After smoothing the objective function by adopting the sigmoid function, we develop an alternating stochastic variance reduced gradient (SVRG) algorithm with a fast convergence rate to solve the problem. 
Simulation results demonstrate the effectiveness of the proposed algorithm and the importance of deploying an RIS in reducing the MSE outage probability.

\end{abstract}

\section{Introduction}
It is envisioned that Internet of Things (IoT) will provide ubiquitous connectivity for billions of low-cost devices with sensing and communication capabilities, thereby enabling automated operations for various intelligent services \cite{IOT}. 
Aggregating data from a large amount of distributed IoT devices is an important but also challenging task. 
The conventional transmit-then-compute approach is not spectrum-efficient and incurs excessive delay in dense IoT networks. 
Fortunately, over-the-air computation (AirComp), integrating communication and computation, has the potential to enable ultra-fast data aggregation by allowing the concurrent data transmission and leveraging the superposition property of multi-access channels \cite{sum1,sum4}.

With great potentials, AirComp has recently attracted considerable research interests. 
In particular, the authors in \cite{info} for the first time showed that the superposition property of multi-access channels can be exploited to compute the nomographic functions. 
Subsequently, the authors in \cite{chen2018over} adopted AirComp to facilitate multiple linear functions computation with antenna arrays, while taking into account the intranode interference of multiple functions. 
The authors in \cite{optimalpolicy} proposed a computation-optimal policy for AirComp systems and studied the ergodic performance under fading channels.
In addition, AirComp was exploited in \cite{yang8761429} to accelerate the convergence rate of federated learning by overcoming the challenge due to limited communication bandwidth. 
According to the aforementioned studies, the performance of AirComp, quantified by the mean-squared error (MSE), is bottlenecked by the unfavorable channel conditions of IoT devices.

Reconfigurable intelligent surface (RIS) has recently been recognized as a promising technology that is capable of overcoming unfavorable channel conditions by reconfiguring the radio propagation environment \cite{zhang,wang2020,sumrate}. 
In particular, an RIS is a man-made flat surface composed of many passive reflecting elements, each of which can independently scatter and shift the phase of the impinging waves\cite{liangyichang}.
The phase-shift pattern generated by all reflecting elements determines the reflection direction of the incident signal, thereby enhancing the signal power at the receiver. 
RIS was exploited in \cite{wubeamse} to enhance the signal power in multiple-input single-output (MISO) systems. 
The authors in \cite{fu2019reconfigurable} adopted RIS to minimize the power consumption of non-orthogonal multiple access (NOMA) networks. 
Moreover, the authors in \cite{jiang2019over} leveraged the benefits of RIS to reduce the MSE of AirComp. 
However, all the aforementioned studies relied on perfect instantaneous channel state information (CSI) to design effective transmit/receive beamforming vectors. 

It is generally difficult to obtain accurate instantaneous CSI in RIS-assisted cellular networks \cite{imperfectcsi,xia}. 
Without perfect CSI, a widely adopted approach is to approximate the actual channel coefficient by the estimated channel coefficient and the channel estimation error, which is assumed to subject to a certain bounded perturbation. 
In this case, robust beamforming can be adopted to address the channel uncertainty\cite{robust}. 
However, the robust beamforming design is very conservative as it only guarantees the worst case performance. 
Besides, an optimal phase shift design of the RIS was proposed in \cite{statistical} that based on the upper bound of the ergodic spectral efficiency.
However, this work relied on the prior knowledge of the underlying channel distribution, which may not be available in many practical scenarios.
     
In this paper, we consider an RIS-aided IoT network, where a multi-antenna access point (AP) aggregates sensing data from multiple IoT devices using AirComp with the assistance of an RIS. 
Without any prior knowledge on the underlying channel distribution, we formulate a stochastic optimization problem that maximizes the probability of MSE being smaller than a certain threshold. 
However, the formulated problem turns out to be a highly intractable non-convex optimization problem, which faces many challenges, including non-smooth probabilistic objective function, coupled optimization variables, and non-convex unimodular constraints. 
To this end, we propose a data-driven approach that relies on the historical channel realizations to jointly optimize the receive beamforming vector at the AP and the phase-shift matrix at the RIS. 
To handle the non-smoothness of the objective function, we adopt a sigmoid function as surrogate, resulting in a continuous optimization problem. 
To decouple the optimization variables and tackle the unimodular constraints, we propose an alternating stochastic variance reduced gradient (SVRG) algorithm with a fast convergence rate. 
Simulation results demonstrate the effectiveness of the proposed algorithm and show the performance gain achieved by deploying an RIS in terms of minimizing the MSE.

\emph{Notations:}
 Matrices, vectors, and scalars are denoted by bold capital, bold lowercase, and lowercase letters, respectively.
  $(\bm \cdot)^{\sf{H}}$ and $(\bm \cdot)^{\sf{T}}$ stand for conjugate transpose and transpose of a matrix or a vector, respectively.
 $|\bm \cdot|$ and $\|\bm \cdot\|$ denote the $l_1$ and $l_2$ norm operators, respectively.
 $\Re[\bm\cdot]$ and $\Im[\bm \cdot]$ represent the real and imaginary parts of a complex matrix, vector, or scalar, respectively.
 $\mathbb{E}\left[\bm \cdot\right]$ and $\rm{Pr}(\bm \cdot)$ denote the expectation of a random variable and probability of an event, respectively.

\section{System Model and Problem Formulation}\label{sys}
\subsection{System Model}
We consider the concurrent uplink transmission via AirComp in an RIS-aided IoT system consisting of $K$ single-antenna IoT devices, an AP with $N$ antennas, and an RIS equipped with $M$ passive reflecting elements. We denote $\mathcal{K}=\{1,2,\ldots, K\}$ as the set of  device indices.
 We denote $z_k\in\mathbb{C}$ as the information-bearing signal at device $k$. 
 The AP aims to recover a nomographic function (e.g., arithmetic mean) of the sensing data from all devices.
  The target function can be expressed as
\setlength\arraycolsep{2pt}
\begin{align}
f=\psi\left(\sum_{k\in\mathcal{K}}\varphi_k(z_k)\right),
\end{align}
where $\varphi_k(\cdot)$ denotes the pre-processing function at device $k$ and $\psi(\cdot)$ denotes the post-processing function at the AP\cite{chen2018uniform}. 
We denote $s_k = \varphi_k(z_k)$ as the transmitted data at device $k$,
and assume that $\{s_k\}$ are independent and have zero mean and unit power, i.e., $\mathbb{E}[s_{k}s_{k}^{\sf H}] = 1$, and $\mathbb{E}[s_{k}s_{j}^{\sf H}] = 0, \forall  k \neq j$.
The target function to be estimated is given by
\begin{align}
	g=\sum_{k\in\mathcal{K}}s_k.
\end{align}
We assume that all IoT devices are synchronized and transmit concurrently to the AP \cite{chen2018over}.
The signal received at the AP from all the devices is given by
\begin{align}
\bm{y}=\sum_{k\in\mathcal{K}}(\bm{h}_{d,k}+\bm G \bm\Theta \bm h_{r,k} ){w}_ks_k+\bm{n},
\end{align}
where $w_k\in\mathbb{C}$ denotes the transmit scalar of device $k$, $\bm{\Theta} = \text{diag}\left\{e^{j\theta_1},\cdots,e^{{j\theta}_M}\right\}$ is the 
 diagonal phase-shift matrix of the RIS with $\theta_m \in [0,2\pi)$, and $ \bm h_{d,k}\in\mathbb{C}^{N\times1}, \bm G\in\mathbb{C}^{N\times M}$, and $\bm h_{r,k}\in\mathbb{C}^{M\times 1}$ are the channel coefficients from device $k$ to the AP, from the RIS to the AP, and from device $k$ to RIS, respectively.
 Besides, $ \bm n\sim\mathcal{CN}(0, \sigma^2\bm I_N) $ is the additive white Gaussian noise (AWGN) with zero mean and power $\sigma^2$. 
Each device has a maximum transmit power, denoted as $P$. 
Hence, we have $|w_k|^2 \leq P, \forall \, k\in\mathcal{K}$.

\subsection{Problem Formulation}
The estimated function before post-processing at the AP is given by
\begin{align}
\hat{g}&={1\over{\sqrt \eta}}{\bm{m}^{\sf{H}}\bm{y}}\! \nonumber \\ 
&=\!{1\over{\sqrt \eta}}{\bm{m}}^{\sf{H}}\sum_{k\in\mathcal{K}}(\bm{h}_{d,k} \!+\!\bm G \bm\Theta \bm h_{r,k} ){w}_k s_k \!+\!{1\over{\sqrt \eta}}\bm{m}^{\sf{H}}\bm{n},
\end{align} 
where $\bm{m}\in\mathbb{C}^N$ denotes the receive beamforming vector and $\eta$ is power normalizing factor. 

We adopt MSE to evaluate the distortion of $\hat{g}$ with respect to $g$, which quantifies the AirComp performance 
\begin{align}
{\sf{MSE}}(\hat{g}, g)&=\mathbb{E}\left(|\hat{g}-g|^2\right)\nonumber \\
&= \!\! \sum_{k\in\mathcal{K}}\left|   \frac{{{\bm{m}}^{\sf{H}}(\bm{h}_{d,k}+\bm G \bm\Theta \bm h_{r,k} ){w}_k}}{\sqrt{\eta}} -1\right|^2 \!+\! \frac{\sigma^2\|\bm{m}\|^2}{\eta}\nonumber . 
\end{align}
To minimize the estimation error, we need to jointly optimize transmit scalar $w_k$, receive beamforming vector $\bm m$, and phase-shift matrix $\bm \Theta$.

When receive beamforming  vector $\bm{m}$ and phase-shift matrix $\bm \Theta$ are given, the optimal transmit scalar that minimizes the MSE can be expressed as \cite{chen2018over,yang8761429,jiang2019over}
\begin{align}\label{a}
w_k^{\star}=\sqrt{\eta}{{(\bm{m}^{\sf{H}}(\bm{h}_{d,k}+\bm G \bm\Theta \bm h_{r,k} ))^{\sf{H}}}\over{\|\bm{m}^{\sf{H}}(\bm{h}_{d,k}+\bm G \bm\Theta \bm h_{r,k} )\|^2}}. 
\end{align}
Due to the transmit power constraint, $\eta$ can be expressed as
\begin{align}\label{b}
\eta=P\min_{k\in\mathcal{K}} \|\bm{m}^{\sf{H}}(\bm{h}_{d,k}+\bm G \bm\Theta \bm h_{r,k} )\|^2.
\end{align}
With \eqref{a} and \eqref{b}, the MSE  can be further rewritten  as
\begin{align}
{\sf{MSE}}(g,\hat g)&={{\|\bm{m}\|^2\sigma^2}\over{\eta}} \nonumber\\ 
&={{\|\bm{m}\|^2\sigma^2}\over{P\min_{k\in\mathcal{K}} \|\bm{m}^{\sf{H}}(\bm{h}_{d,k}+\bm G \bm\Theta \bm h_{r,k} )\|^2}}.\nonumber
\end{align}

To minimize the MSE, the instantaneous CSI is required, which however is
impratical to obtain\cite{xia}. 
In this paper, we completely forego the assumption of the availability of instantaneous CSI.
 We maximize the probability that the MSE falls below to a predetermined threshold denoted as $\tau$.
Hence, the formulated problem is given by 
\begin{align} \label{p0}
\mathop{\rm{maximize}}_{\bm{m},\bm \Theta}& \quad{\rm{Pr}}\left({\sf{MSE}}(g,\hat g) \leq \tau\right) \nonumber \\
\text{subject to} &\quad 0\leq \theta_m \leq 2\pi, m=1,\ldots,M.
\end{align}
We define $f(\bm m, \bm v)$ as the MSE outage probability, which is given by 
\begin{align}
	f(\bm m, \bm v):=1-{\rm{Pr}}\left({\sf{MSE}}(g,\hat g) \leq \tau\right).
\end{align}

\noindent By denoting  $\gamma = \tau P/\sigma^2$,
  problem \eqref{p0} can be equivalently expressed as
\begin{align}\label{p1}
\mathop{\rm{maximize}}_{\bm{m},\bm \Theta}& \quad {\rm{Pr}}\left(\max_{k\in\mathcal{K}} {{\|\bm{m}\|^2}\over{\|\bm{m}^{\sf{H}}(\bm{h}_{d,k}+\bm G \text{diag}(\bm h_{r,k})\bm v )\|^2}} \leq \gamma\right)\nonumber \\
\text{subject to}& \quad |v_m|=1,  m=1,\ldots,M, 
\end{align}
\noindent where $ \bm v=[v_1,\ldots,v_M] ^{\sf T}=[\mathrm{e}^{j\theta_1},\ldots, \mathrm{e}^{j\theta_M}]^{\sf H}$.

 To facilitate problem transformation, we define 
\begin{align}\label{obj}
d(\bm m,\bm v;\bm h_{e,k}  )&:= \|\bm{m}\|^2 - \gamma\|\bm{m}^{\sf{H}}(\bm{h}_{d,k}+\bm G \text{diag}(\bm h_{r,k})\bm v )\|^2, \nonumber 
\end{align} 
where $ \bm h_{e,k} = \{\bm h_{d,k},\bm h_{r,k},\bm G \} $  is an abstraction of channel coefficient between device $k$ and the AP. Thus, for all $k\in \mathcal{K}$, we have
\begin{align}
{{\|\bm{m}\|^2}\over{\|\bm{m}^{\sf{H}}(\bm{h}_{d,k}+\bm G \text{diag}(\bm h_{r,k})\bm v )\|^2}} \leq \gamma\Leftrightarrow 
d(\bm m,\bm v;\bm h_{e,k}) \leq 0.\nonumber
\end{align} 
As a result, the MSE outage probability can be rewritten as 
\begin{align}
& f(\bm m, \bm v) \nonumber \\ 
&= 1- {\rm{Pr}}\left({{\|\bm{m}\|^2}\over{\|\bm{m}^{\sf{H}}(\bm{h}_{d,k}+\bm G \text{diag}(\bm h_{r,k})\bm v)\|^2}} \leq \gamma,\forall k\in\mathcal{K}\right) \nonumber\\&= \operatorname{Pr}\left\{\Big(\mathop{\text{max}}_{ k\in\mathcal{K}}d(\bm m, \bm v;\bm h_{e,k})\Big)>0\right\} \nonumber \\
&= \mathbb{E}\left[\mathcal{I}_{(0,+\infty)}\Big(\mathop{\text{max}}_{ k\in\mathcal{K}}d(\bm m, \bm v; \bm h_{e,k})\Big)\right],
\end{align} 
where $\mathcal{I}_{(0,+\infty)}(x) $ is an indicator function defined as 
\begin{align}\label{indicator}
\mathcal{I}_{(0,+\infty)} (x)=\begin{cases}
1,\quad\text{if}\quad x>0,\\
0,\quad\text{otherwise}.
\end{cases}
\end{align}
The channel distribution is indispensable when it comes to compute $f(\bm m, \bm v)$ accurately.
Due to the ergodicity of the channel process, we can approximate $f(\bm m, \bm v)$ by averaging over a set of historical channel realizations, i.e., sample average. 
The approximation converges when the number of historical channel samples is sufficiently large. 
In particular, we denote the channel sample set as $\mathcal{H}_T = \{ \bm h_{e,k}^t \}_{t=1}^{T}$ with $T$ samples, where $\bm h_{e,k}^t$ denotes the $t$-th channel sample of device $k$.
We assume that these historical channel samples are available at the AP based on the previous measurements, as in  \cite{shiyunmei}.
With a set of channel samples, we adopt the following sample average to approximate $f(\bm m, \bm v)$ 
\begin{align}
	\bar f(\bm m, \bm v;\mathcal{H}_T)=\frac{1}{T}\sum_{t=1}^{T} \mathcal{I}_{(0,+\infty)}\Big(\mathop{\text{max}}_{ k\in\mathcal{K}}d(\bm m, \bm v; \bm h_{e,k}^t)\Big).
\end{align} 	
As a result, we rewrite problem \eqref{p1} as 
\begin{align}\label{p2}
\mathop{\text{minimize}}_{\bm{m} ,\bm v}  \quad &\bar f(\bm m, \bm v;\mathcal{H}_T) \nonumber \\
\text{subject to}  \quad & |v_m|=1, m=1,\ldots,M.
\end{align}
Problem (\ref{p2}) is an MSE outage probability minimization problem, which is still highly intractable due to the following challenges. First, the indicator function is non-convex and discontinuous.
Second, the unimodular constraint of the phase shift of each RIS element is not convex.
Third, the optimization variables $\bm m$ and $\bm v$ are coupled in the objective function.

\section{Problem Decomposition} \label{dec}

\subsection{Sigmoid Function for Indicator Function}
To deal with the discontinuiality of the indicator function $\mathcal{I}_{(0,+\infty)}(x)$, we adopt the smooth sigmoid function as its surrogate \cite{shiyunmei}
\begin{align}\label{sigmoid}
S(x) = {1}/{(1+e^{-x})},
\end{align}
where $x\in \mathbb{R}$. Although the sigmoid function is not convex, it is continuously differentiable and strictly monotonic increasing.  
With the smooth surrogate, problem \eqref{p2} is given by 
\begin{align}\label{p3}
\mathop{\text{minimize}}_{\bm{m} ,\bm v}&  \quad \frac{1}{T}\sum_{t=1}^{T} S\Big(\mathop{\text{max}}_{ k\in\mathcal{K}}d(\bm m, \bm v; \bm h_{e,k})\Big), \nonumber \\
\text{subject to}  &\quad   |v_m|=1, m=1,\ldots,M.
\end{align}
\noindent Problem \eqref{p3} is a continuous stochastic optimization problem.

The coupling between optimization variables $\bm m$ and $\bm v$ makes the problem difficult to be solved.
In the following two subsections, we shall decouple the variables by alternatively designing the receive beamforming vector and the phase shift vector.

\subsection{Receive Beamforming Vector Design}
For fixed $ \bm v $, we denote $\bm h_k= \bm{h}_{d,k}+\bm G \text{diag}(\bm h_{r,k})\bm v  \in \mathbb{C}^{N \times 1}$, and rewrite $d(\bm m,\bm v;\bm h_{e,k}  )$ as 
\begin{equation}
	d(\bm m;\bm h_k) = \|\bm{m}\|^2 - \gamma\|\bm h_k^{\sf H}\bm m\|^2. \nonumber
\end{equation}
Therefore, problem $ \eqref{p3} $ can be simplified as 
\begin{align}\label{p4}
\mathop{\text{minimize}}_{\bm m} \quad \frac{1}{T}\sum_{t=1}^{T} S\Big(\mathop{\text{max}}_{ k\in\mathcal{K}}d(\bm m; \bm h_k^t)\Big).
\end{align}
It is worth noting that $\bm h_k^t$ is obtained based on $\bm h_{e,k}^t$.
To facilitate efficient algorithm design, we define 
\begin{align}
\tilde{\bm m}&:=[\Re[\bm m]^{\sf T}, \Im[\bm m]^{\sf T}]^{\sf T}\in \mathbb{R}^{2N},\\  
\tilde{\bm H}_k&:= \begin{bmatrix}
\Re[\bm h_k] & -\Im[\bm h_k] \\
\Im[\bm h_k] & \Re[\bm h_k]
\end{bmatrix}\in \mathbb{R}^{2N\times2}.
\end{align}
\noindent We can equivalently express problem$\eqref{p4}$ in terms of real variables (i.e., $\tilde{\bm m}$ and $\tilde{\bm H}_k$) as
\begin{align}\label{subm}
\mathop{\text{minimize}}_{\tilde{\bm m}} \quad u_1(\tilde{\bm m})=\frac{1}{T}\sum_{t=1}^{T}S\Big(\mathop{\text{max}}_{k\in\mathcal{K}}d(\tilde{\bm m};\tilde{\bm H}_k^t)\Big),
\end{align}
where $d(\tilde{\bm m};\tilde{\bm H}_k) = \|\tilde{\bm m}\|^2 - \gamma\|\tilde{\bm H}_k^{\sf T}\tilde{\bm m}\|^2$.

\subsection{Phase-Shift Vector Design}
On the other hand, for fixed $ \bm m $, we denote $b_k = \bm m^{\sf H} \bm h_{d,k} \in \mathbb{ C}$ and $ \bm a_{k}^{\sf H} =\bm m^{\sf H}\bm G \text{diag}(\bm h_{r,k}) \in \mathbb{ C}^{1\times M}$. Therefore, we rewrite $d(\bm m,\bm v;\bm h_{e,k}  )$ as
\begin{align}\label{fixm}
d(\bm v; \bm a_k,b_k) = \|\bm{m}\|^2 - \gamma\|b_k+\bm a_{k}^{\sf H}\bm v\|^2,
\end{align} 
where  $\bm a_{k}^{\sf H}\bm v  =  \bm m^{\sf H}\bm G \text{diag}(\bm h_{r,k}) \bm v $.
By defining 
\begin{align}
\tilde{\bm v}&:=[\Re[\bm v]^{\sf T}, \Im[\bm v]^{\sf T}]^{\sf T}\in \mathbb{R}^{2M},\\
\tilde{\bm b}_k&:=[\Re[b_k], \Im[ b_k] ]^{\sf T}\in \mathbb{R}^{2},\\
\tilde{\bm A}_k&:= \begin{bmatrix}
\Re[\bm a_k] & -\Im[\bm a_k] \\
\Im[\bm a_k] & \Re[\bm a_k]
\end{bmatrix}\in \mathbb{R}^{2M\times2},
\end{align} 
we rewrite $d(\bm v; \bm a_k,b_k)$ as
\begin{align}
d(\tilde{\bm v};\tilde{\bm A}_k,\tilde{\bm b}_k) = \|\tilde{\bm m}\|^2 - \gamma\| \tilde{ \bm b_{k}}+\tilde{\bm A}_k^{\sf T}\tilde{\bm v}\|^2.
\end{align}
We equivalently express problem$\eqref{p3}$ in terms of real variables (i.e., $\tilde{\bm v}$, $\tilde{\bm b}_k$, and $\tilde{\bm A}_k$) as
\begin{align}\label{subv}
\mathop{\text{minimize}}_{\tilde{\bm v}}& \quad  u_2(\tilde{\bm v})=\frac{1}{T}\sum_{t=1}^{T} S\Big(\mathop{\text{max}}_{ k\in\mathcal{K}}d(\tilde{\bm v};\tilde{\bm A}_k^t,\tilde{\bm b}_k^t)\Big)\nonumber\\
\text{subject to}& \quad \vert \tilde{v}_m\vert^2 + \vert \tilde{v}_{m+M}\vert^2=1, m = 1,\ldots,M.
\end{align}

Problems (\ref{subm}) and (\ref{subv}) are still non-convex. 
We shall develop an efficient alternating SVRG algorithm to solve both problems in the following section.

 
\section{Proposed Alternating SVRG Algorithm} \label{alt}

The objective functions of problems \eqref{subm} and \eqref{subv} (i.e., $u_1(\tilde{\bm m})$ and $u_2(\tilde{\bm v})$) are continuous and differentiable. 
Stochastic gradient descent (SGD) has been extensively applied to solve this kind of finite-sum-form problems \cite{SGD}.
However, inherent variance is inevitable, as SGD approximates the full gradient by a single gradient.
To ensure convergence, the step size has to be decayed to zero.
Thus, the SGD method suffers from slow convergence. 
Fortunately, the SVRG algorithm can effectively address this issue \cite{SVRG}.
Although there is some loss in complexity, SVRG achieves a faster convergence rate than SGD.
Hence, we propose an alternating SVRG algorithm to solve problems \eqref{subm} and \eqref{subv} alternatively until the convergence. 
Moreover, the mini-batch version of SVRG is adopted to further reduce the variance and enhance the parallelism\cite{batch}.

\subsection{Receive Beamforming Vector Optimization}
In this subsection, we update $\tilde {\bm m}$ by applying SVRG to minimize $u_1(\tilde{\bm m})$.
Different from SGD, each epoch of SVRG has an inner loop. 
We denote $R$ and $Q$ as the number of epochs and the number of iterations in each epoch, respectively. 
Specifically, we randomly choose mini-batch samples $\mathcal{I}_q\subset\{1,\cdots,T\}$ for the $q$-th iteration in the inner loop and then update $\tilde{\bm m}$ according to the following rule 
\begin{align}
\tilde{\bm m}_{q+1}^{r+1} = \tilde{\bm m}_{q}^{r+1} - \alpha_{\tilde{m}}\bm{g}_{q}^{r+1},
\label{update_m}
\end{align}
where $\alpha_{\tilde{m}}$ is the step size and $\bm g_{q}^{r+1}$ is the descent direction of the $(rQ+q)$-th iteration (i.e., $q$-th iteration at the $(r+1)$-th epoch). 
In addition, $\bm{g}_{q}^{r+1}$ can be calculated by 
\begin{align} \label{grad_m}
\bm{g}_{q}^{r+1} =& {1 \over {|\mathcal{I}_q|}}\sum_{i\in \mathcal{I}_q}\Big(\nabla S\big(\mathop{\text{max}}_{k\in\mathcal{K}}d(\tilde{\bm m}^{r+1}_q;\tilde{\bm H}_k^i)\big)  \nonumber \\   &-\nabla S\big(\mathop{\text{max}}_{k\in\mathcal{K}}d(\tilde{\bm m}^r_Q;\tilde{\bm H}_k^i)\big) \Big )+ \nabla u_1(\tilde{\bm m}^r_Q), 
\end{align}
where $|\mathcal{I}_q|$ denotes the cardinality of $\mathcal{I}_q$ and $\nabla u_1(\tilde{\bm m}^r_Q)$ is the batch gradient computed at the end of the $rQ$-th iteration and used in the next epoch's $Q$ iterations. 

The gradient of $S\big(\mathop{\text{max}}_{k\in\mathcal{K}}d(\tilde{\bm m};\tilde{\bm H}_k)\big)$ is presented in the following lemma. 
\begin{lemma}
  The gradient of  $S\big(\mathop{\text{max}}_{k\in\mathcal{K}}d(\tilde{\bm m};\tilde{\bm H}_k)\big)$ with respect to $\tilde{\bm m}$ denoted as $\nabla_{\tilde{\bm m}}S\big(\mathop{\text{max}}_{k\in\mathcal{K}}d(\tilde{\bm m};\tilde{\bm H}_k)\big)$ is given by
  \begin{align} \label{Eq_L1}
  \nabla_{\tilde{\bm m}}S\big(\mathop{\text{max}}_{k\in\mathcal{K}}d(\tilde{\bm m};\tilde{\bm H}_k)\big) & = 
	 S\big(d(\tilde{\bm m};\tilde{\bm H}_{k^*}^t)\big) \nonumber \\
	 & \hspace{-25mm} \times (1-S\big(d(\tilde{\bm m};\tilde{\bm H}_{k^*}^t)\big)\big)(2\tilde{\bm m} - 2\gamma\tilde{\bm H}_{k^*} \tilde{\bm H}_{k^*}^{\sf T}\tilde{\bm m}), 
  \end{align} 	
  $\textrm{where}\  k^* = \arg \mathop{\text{max}}_{k\in\mathcal{K}} d(\tilde{\bm m};\tilde{\bm H}_k^t).$
\end{lemma}

By substituting (\ref{Eq_L1}) into (\ref{grad_m}), we obtain $\bm{g}_{q}^{r+1}$, which can in turn be used to update $\tilde{\bm m}$ according to (\ref{update_m}).

\subsection{Phase-Shift Vector Optimization}
Similarly, we update the phase-shift vector by applying SVRG as follows
\begin{align} \label{update_v}
\tilde{\bm y}^{r+1}_{q+1} &= \tilde{\bm v}^{r+1}_{q} - \alpha_{\tilde{v}}\nabla_{\tilde{\bm v}}\bm g_q^{r+1}, \\
\label{update_g1}\bm{g}_q^{r+1} &= {1 \over {|\mathcal{I}_q|}}\sum_{i\in \mathcal{I}_q}\Big(\nabla S\big(\mathop{\text{max}}_{k\in\mathcal{K}}d(\tilde{\bm v}_q^{r+1};\tilde{\bm A}_k^i,\tilde{\bm b}_k^i)\big)  \nonumber \\   & \hspace{3mm}-\nabla S\big(\mathop{\text{max}}_{k\in\mathcal{K}}d(\tilde{\bm v}_Q^{r};\tilde{\bm A}_k^i,\tilde{\bm b}_k^i)\big) \Big )+ \nabla u_2(\tilde{\bm v}^r_Q), \\
(\tilde{v}^{r+1}_{q+1})_m & \!=\! \frac{(\tilde{y}^{r+1}_{q+1})_m}{(\vert(\tilde{y}^{r+1}_{q+1})_m\vert^2+\vert(\tilde{y}^{r+1}_{q+1})_{m+M}\vert^2)^{\frac{1}{2}}}, m=1,\ldots,2M, \label{solve_v}
\end{align}
where $\alpha_{\tilde{v}}$ denotes the update step size of $\tilde{\bm v}$ and $(\bm \cdot)_m$ represents the $m$-th component of a vector.
Due to the unimodular constraint, we take an extra Euclidean projection on $\tilde{\bm y}^{r+1}_{q+1}$ to obtain $\tilde{\bm v}^{r+1}_{q+1}$ according to \eqref{solve_v}, where
$(\tilde{y}^{r+1}_{q+1})_m=(\tilde{y}^{r+1}_{q+1})_{m-2M}$ if $m>2M$.

The gradient of $S\big(\mathop{\text{max}}_{k\in\mathcal{K}}d(\tilde{\bm v};\tilde{\bm A}_k,\tilde{\bm b}_k)\big)$ is provided by the following lemma. 
\begin{lemma} 
	The gradient of  $S\big(\mathop{\text{max}}_{k\in\mathcal{K}}d(\tilde{\bm v};\tilde{\bm A}_k,\tilde{\bm b}_k)\big)$ with respect to $\tilde{\bm v}$ denoted as $\nabla_{\tilde{\bm v}}S\big(\mathop{\text{max}}_{k\in\mathcal{K}}d(\tilde{\bm v};\tilde{\bm A}_k,\tilde{\bm b}_k)\big)$ is given by
	\begin{align} \label{Eq_L2}
	& \nabla_{\tilde{\bm v}}S\big(\mathop{\text{max}}_{k\in\mathcal{K}}d(\tilde{\bm v};\tilde{\bm A}_k,\tilde{\bm b}_k)\big) = S\big(d(\tilde{\bm v};\tilde{\bm A}_{k^*},\tilde{\bm b}_{k^*})\big) \nonumber \\
	& \times \big(1-S\big(d(\tilde{\bm v};\tilde{\bm A}_{k^*},\tilde{\bm b}_{k^*})\big)\big) (-2\gamma\tilde{\bm A}_{k^*}\tilde{\bm b}_{k^*}-  2\gamma\tilde{\bm A}_{k^*} \tilde{\bm A}_{k^*}^{\sf T}\tilde{\bm v}), 
	\end{align}
	$\textrm{where}\  k^* = \arg \max_{k\in\mathcal{K}} d(\tilde{\bm v};\tilde{\bm A}_k,\tilde{\bm b}_k).$ 	
\end{lemma}

By substituting (\ref{Eq_L2}) into (\ref{update_g1}), we obtain $\bm{g}_{q}^{r+1}$, which can then be used to update $\bm{\tilde{v}}$ according to \eqref{update_v} and \eqref{solve_v}.  

The overall algorithm for solving problem \eqref{p3} is summarized in Algorithm \ref{algo1}, where 
$L$ is the number of the iterations that alternately optimize the two optimization variables. 
   
\begin{algorithm} [ht]
	\caption{Alternating SVRG Algorithm}
	\label{algo1}
	\SetKwData{Index}{Index}
    \KwIn{Number of iterations $L,R,Q$.}
    	\For{$l=0,1,\ldots,L-1$}{
    	\For{$r=0,1,\ldots, R-1$}{
    	$\tilde{\bm m}_{0}^{r+1} \leftarrow \tilde{\bm m}_{Q}^{r}$ and compute  $\nabla u_1(\tilde{\bm m}^r_Q)$.\\
    			 \For{$q=0,1,\ldots, Q-1$}{
    			 Choose $\mathcal{I}_q$ uniformly from $\{1,\ldots,T\}$.\\
		 	Fix $\tilde{\bm v}$, update $\tilde{\bm m}$ via \eqref{update_m}. \\
					   			 }}
					   			 \For{$r=0,1,\ldots, R-1$}{
					   			 $\tilde{\bm v}^{r+1}_{0} \leftarrow \tilde{\bm v}^{r}_{Q}$ and compute  $\nabla u_2(\tilde{\bm v}^r_Q)$.\\
					   			          \For{$q=0,1,\ldots, Q-1$}{
					   			          Choose $\mathcal{I}_q$ uniformly from $\{1,\ldots,T\}$.\\
 	Fix $\tilde{\bm m}$, update $\bm{\tilde{v}}$ via \eqref{update_v} and \eqref{solve_v}.\\
						  			  
             }}
   }
\end{algorithm}
%
\section{Simulation Results} \label{sim}
 In this section, we present the simulation results of the proposed algorithm for AirComp in RIS-aided IoT networks. 
 We consider a three-dimentional setting, where the AP and the RIS are located at $(0,0,10)$ and $(20,10,10)$ meters, respectively.
Besides, $K=20$ devices are randomly located in the square centered at $(25,5,0)$ with side length $10$ meters.
In simulations, we consider Rician fading with a factor of 3 for the reflecting link and Rayleigh fading for the direct link to generate historical channel realizations. 
Note that the proposed algorithm can be applied to any other fading models. 
The path loss of a link with length $d$ is modeled as $L(d) = L_0d^{-\beta}$, where $\beta$ denotes the path loss exponent and $L_0 = 10^{-3}$. 
For the device-AP link, the RIS-AP link, and the device-RIS link, the path loss exponents are set to be $3.8$, $2.2$, and $2.2$, respectively. 
We set the maximum transmit power of the AP and the noise power as $P=0$ dBm and $\sigma^2=-100$ dBm, respectively. 
We set the size of the historical channel set as $T = 300$. 
The step sizes are set to $ \alpha_{\tilde{w}}=0.1$ and $ \alpha_{\tilde{v}}=0.01$. 
Besides, we set $L=100$, $R=200$, and $Q=25$. 
The size of the mini-batch is set to be $50$.
All results in the simulations are obtained by averaging over $100$ Monte Carlo realizations. 
 \begin{figure}[t]
  \centering
  \centerline{\includegraphics[width=8cm]{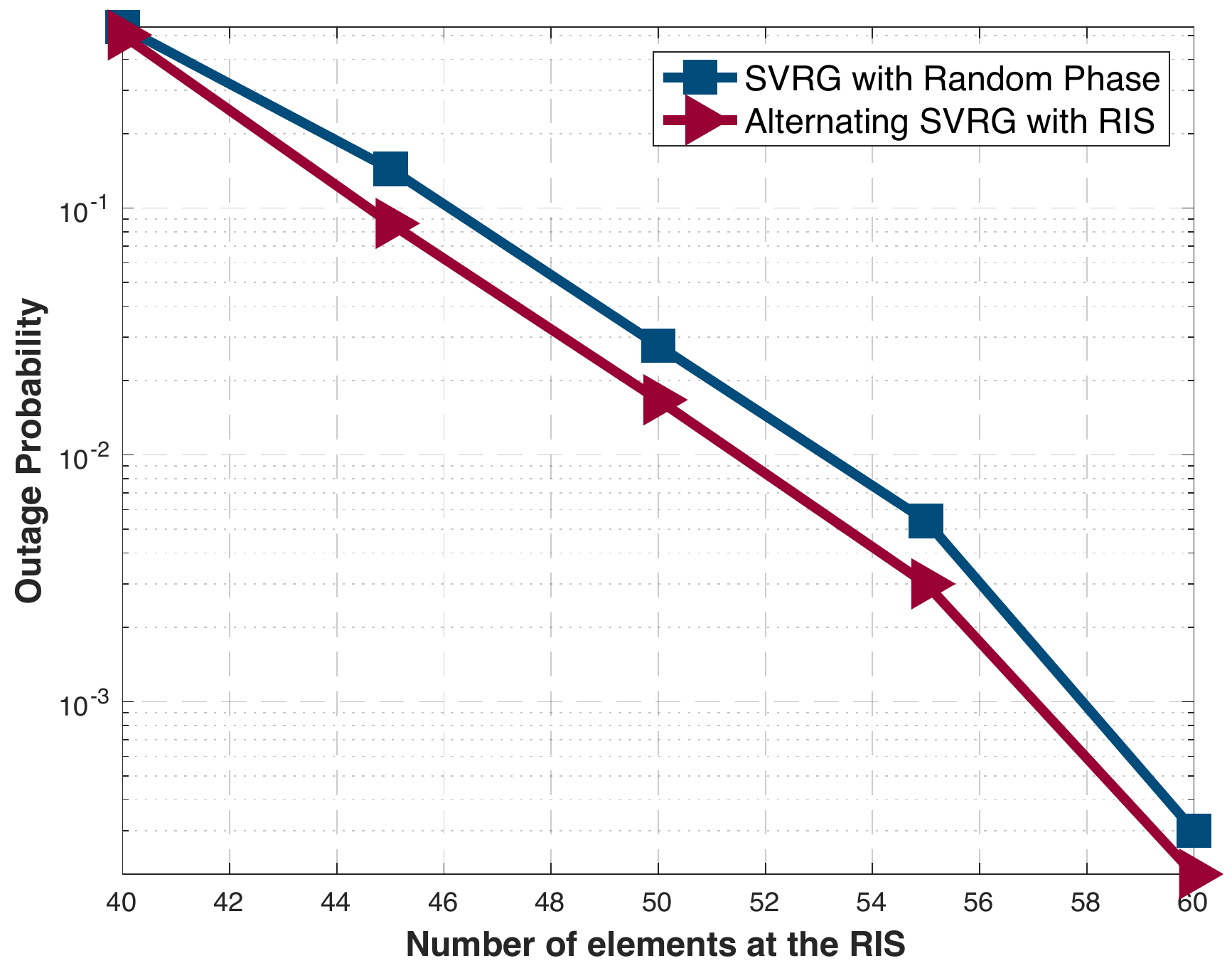}}
  \vspace{-0.2cm}
 	\caption{MSE outage probability versus number of reflecting elements at RIS when $N=20$ and $\tau = -28$ dB.}
	\label{N}
\end{figure}
\begin{figure}[t]
  \centering
  \centerline{\includegraphics[width=8cm]{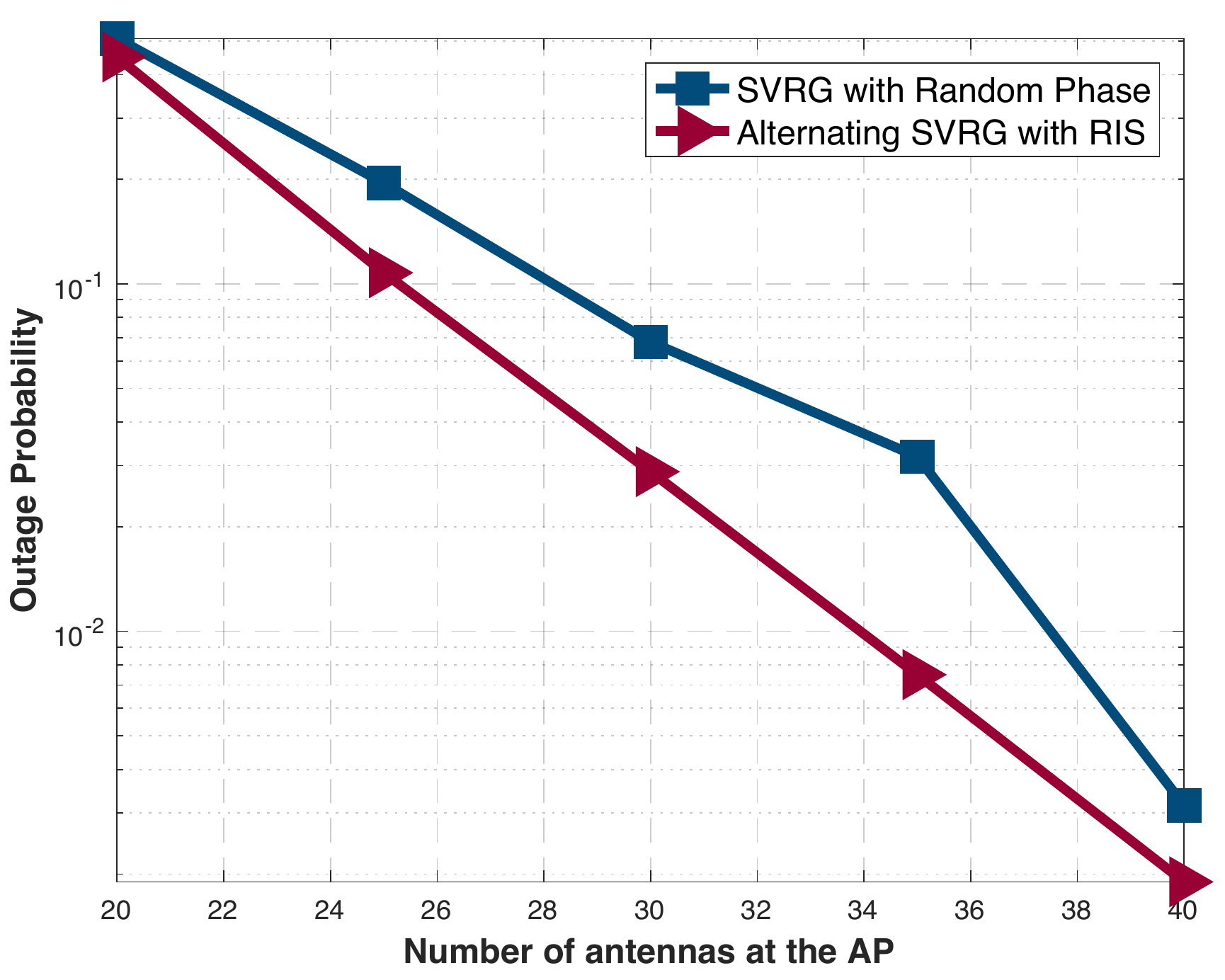}}
  \vspace{-0.2cm}
 	\caption{MSE outage probability versus number of antennas at the AP when $M=40$ and $\tau = -28$ dB.}
	\label{M}
\end{figure}
\begin{figure}[t]
  \centering
  \centerline{\includegraphics[width=8cm]{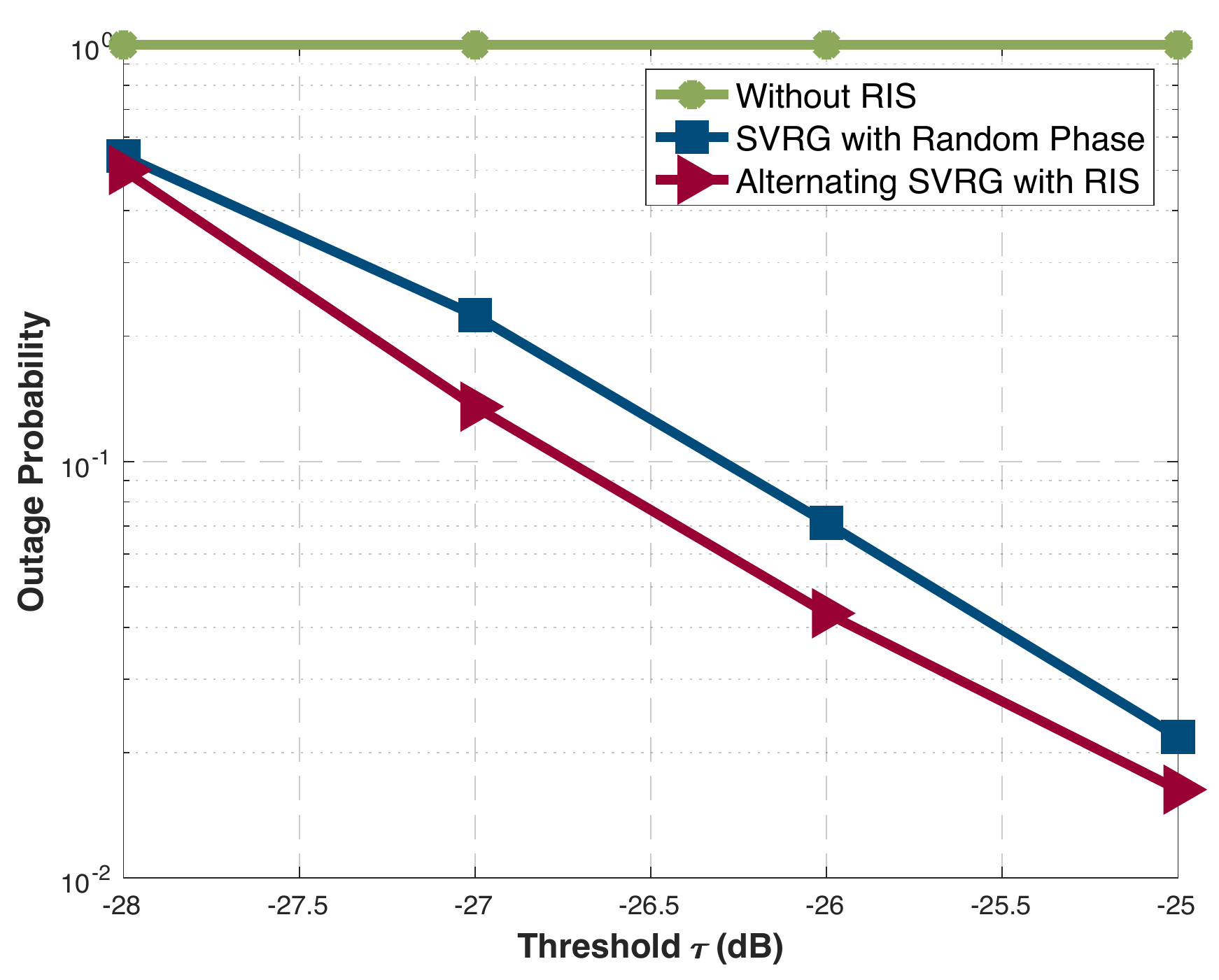}}
  \vspace{-0.2cm}
 	\caption{MSE outage probability versus reception threshold when  $N=20$ and $M=40$.}
	\label{gamma}
\end{figure}

Fig. \ref{N} shows the MSE outage probability versus the number of reflecting elements when $N=20$ and $\tau=-28$ dB. The MSE outage probability decreases as the number of reflecting elements increases. 
This is because an RIS with more elements can better overcome the unfavorable channel conditions and achieve a higher received signal power, which in turn reduce the MSE outage probability of AirComp. 

Fig. \ref{M} plots the MSE outage probability versus the number of antennas at the AP when $M=40$ and $\tau=-28$ dB. 
As the number of antennas at the AP increases, the MSE outage probability decreases for both schemes under consideration. 
This is because, with more receiving antennas, a higher power gain can be achieved to mitigate the detrimental effect of noise, which in turn reduces the MSE outage probability. 
In addition, the proposed algorithm achieves a lower MSE outage probability than the SVRG with random phase scheme, which highlights the importance of optimizing the phase shifts and shows that the proposed algorithm can effectively solve the stochastic optimization problem.

Fig. \ref{gamma} illustrates the MSE outage probability versus threshold $\tau$ when $N=20$ and $M=40$. 
It can be observed that, with the increase of threshold $\tau$, the probability that the achievable MSE is below $\tau$ decreases, which in turn decreases the MSE outage probability. 
Besides, we can also observe that deploying an RIS can significantly reduces the MSE outage probability, as RIS is capable of overcoming the unfavorable channel conditions, thereby eliminating the bottleneck of AirComp.


\section{Conclusions} \label{con}
In this paper, we investigated the MSE outage probability of AirComp in an IoT network with the assistance of an RIS. 
Without instantaneous CSI and any prior knowledge on the underlying channel distribution, we proposed a data-driven approach to jointly optimize the receive beamforming vector at the AP and the phase-shift matrix at the RIS, taking into account the maximum transmit power constraint and the unit modulus constraint. 
The sigmoid function was adopted to smooth the indicator function. 
We developed an alternating SVGR algorithm with a fast convergence rate to solve the original problem. 
Simulation results showed that deploying an RIS can significantly reduce the MSE outage probability by eliminating the bottleneck of AirComp and demonstrated the effectiveness of the proposed algorithm. 

\bibliographystyle{IEEEtran}
\bibliography{ref} 

\end{document}